\pdfoutput=1
\documentclass[a4paper]{article}
\usepackage{geometry}   
\date{} 
\usepackage[numbers]{natbib}   
\usepackage{subcaption}
\usepackage{authblk}
\usepackage [latin1]{inputenc}
\usepackage{cite}
\usepackage{graphicx}	
\graphicspath{ {mypaper}}
\usepackage{subfloat}			
\usepackage{amssymb}
\usepackage[colorlinks,allcolors=blue]{hyperref}
\usepackage{amsmath}
\usepackage{ wasysym }
\usepackage{ upgreek }
\usepackage{float}
\title{Bayes' Theorem, Inflation, and the Ekpyrotic Universe}
\author[1]{Joseph Wraga\footnote{\href{mailto:jmw465@drexel.edu}{jmw465@drexel.edu}}}
\author[1]{David M. Goldberg}
\affil[1]{\textit{Department of Physics, Drexel University, Philadelphia, PA 19104}}

\begin{document}
\maketitle

\begin{abstract}
We calculate the Bayesian evidences for a class of Ekpyrotic universe models, and compare with a model of single field inflation with a Higgs-type potential. Combining parsimony and observational constraints, this gives us a systematic way to evaluate the degree to which Ekpyrotic models are constrained by CMB data from Planck. We integrate the equations of motion numerically to define a likelihood using Planck 2018 data and sample this likelihood to obtain Bayesian evidences. Priors are justified and used to put Ekpyrotic models and inflation on equal footing. We find reasonable preference for one of the considered Ekpyrotic models over the others, but that even this one is disfavored compared with Higgs inflation.
\end{abstract}

\section{Introduction}

The theory of cosmic inflation \citep{guth, linde, starobinsky, mukh, albrecht,planck2018param,planck2018NG} is the current leading paradigm for the origin of cosmic structure. Inflation provides a mechanism for the observed flat, isotropic universe with a nearly scale-invariant and gaussian spectrum of perturbations, as observed in temperature and polarization fluctuations in the CMB by the Planck telescope \citep{planck2018param,planck2018NG,planck2018inf}, combined with Baryon Acoustic Oscillation data \citep{wmap, sdss}, Large Scale Structure data \citep{ross, ly}, and type Ia supernova observations \citep{sn}.  

While textbooks often refer to this as ``the'' theory of inflation, there are many possible inflationary models which lead to predictions compatible with Planck \citep{best}. These models are defined by the form of the inflationary potential $V(\phi)$ in a theory defined by the action for a scalar field $\phi$ minimally coupled to gravity
\begin{align} S = \int d^4x \sqrt{-g}\left[\frac{R}{2}+\frac{1}{2}\partial_{\mu} \phi \partial^{\mu} \phi-V(\phi) \right], \end{align}
where $R$ is the Ricci scalar curvature associated with an expanding, nearly-de-Sitter spacetime with $(+,-,-,-)$ metric signature and $g$ is the metric determinant. We use units in which $M_{Pl}^{-2} = 8\pi G = 1 $. 

Given an inflationary potential and initial conditions, we can calculate the predictions for the resulting cosmological observables (as well as the temperature and polarization multipoles). 
However, there are many possibilities for the inflationary potential proposed in the literature- well over a hundred of even single-field ones- varying greatly in complexity. So a question that arises is how can we prefer one over another? 

All else being equal, we would prefer a model with fewer parameters which fits the data. Several authors \citep{hunting,assess,sky,param,best} have attempted to describe the posterior probability space with Bayes' theorem, which incorporates an Occam's razor that favors simple models that fit data as well as complex ones. These authors have used the Planck constraints on the observables to constrain the possible parameter space and perform model selection using the ratio of Bayesian evidences (integrated likelihoods). In this paper, we propose a similar approach for a broader category of early universe models to answer our central question: Given that simple inflation models fit the data well, do we have reason to consider seriously a different class of models which require more model parameters?

Alternatives to inflation include theories in which density perturbations are generated during a contracting phase preceding a bounce$\textendash$ a reversal from contraction to expansion $\textendash$ after which the perturbations are imprinted on the CMB. One such scenario is the Ekpyrotic theory, in which the universe contracts slowly driven by a scalar field $\phi$ characterized by high pressure \citep{after,colliding,ek}. Originally formulated in terms of string theory, many recent Ekpyrotic scenarios (which we focus on here) are low-energy effective theories based on scalar field theory embedded in a $(3+1)-$dimensional spacetime. These theories require a second scalar field $\chi$ so that entropy perturbations are generated, and these are then converted into the curvature perturbation responsible for the CMB observables (e.g. the power spectrum and bispectrum) at a later phase due to a repulsive potential which causes a bend in scalar field space.

There are various inflationary and Ekpyrotic models proposed in the literature. There are philosophical reasons to prefer one over the other, for example (where $+,-$ means in favor of or against Ekpyrosis and a bullet point means it is inconclusive):
\begin{itemize}
  \item[$+$] The possibility of a multiverse in eternal inflation and the possible effects of Planck level physics near the time of the initial singularity \citep{planck2013}. These issues are avoided in this class of Ekpyrotic models.
  \item[$-$] Details of a bounce phase \citep{nonmin} are speculative.
  \item The reheating process in either case is largely unknown \citep{reheat}.
  \item The presence or absence of primordial gravitational waves \citep{planck2013, quest}; inflation models predict some level (though varying by orders of magnitude) whereas Ekpyrosis generically leads to a negligibly small amplitude. 
\end{itemize}
Here, we consider what it would take for us to prefer Ekpyrosis over inflation based only on the primordial observables from Planck.


Many Bayesian analyses have been conducted for many inflationary models \citep{param,best}. To our knowledge, Bayesian parameter estimation and model selection has not been done for Ekpyrotic models. These have more parameters than the simplest inflation models, so the question is whether these extra parameters are supported by the data, or if the added complexity makes these disfavorable compared with inflation. 

The outline of this paper is as follows. In $\S$ 2, we describe the models of interest and how they lead to predictions for the cosmological  observables. In $\S$ 3, we discuss our pipeline for calculating these observables given the input parameters, calculating the likelihoods as functions of model parameters, Bayesian model comparison, and our choice of priors. In $\S$ 4 we present and discuss our results. 

\section{Ekpyrotic models and Observables}
The Ekpyrotic theory was originally proposed via colliding branes in string theory \citep{colliding}. The idea is that the universe undergoes a cyclic process wherein a hot big bang phase is triggered by the collision of the two branes. Some recent Ekpyrotic model building is based in a $(3+1)-D$ effective theory, ignoring string theory effects and a singularity. We focus on these types of models.

In an Ekpyrotic universe, the big bang was preceded by a phase of mild contraction of the scale factor due to a high pressure fluid (modeled by a scalar field), which solves the cosmological problems (the flatness and horizon problems) as inflation does \citep{colliding, review}.  The Ekpyrotic-bounce model we consider is one of two scalar fields minimally coupled to gravity defined by the action \citep{after}
\begin{align} S = \int d^4x \sqrt{-g}\left[\frac{R}{2}+P(X,\phi)-\frac{1}{2}\Omega^2(\phi)(\partial \chi)^2-V(\phi,\chi) \right] \end{align}
where $R$ is the Ricci scalar curvature associated with a slowly contracting, nearly-Minkowski spacetime with $(+,-,-,-)$ metric signature. We will discuss in more detail below, but for most of cosmic evolution, $P(X,\phi) \approx X \equiv \frac{1}{2}\partial_{\mu} \phi \partial^{\mu} \phi$, as is the case in inflation, eq. (1). $P(X,\phi)$ will only have its more complex form during the bounce. The field $\phi$ will drive both Ekpyrosis and the nonsingular bounce, and we discuss the form of this potential below. The field $\chi$ is orthogonal in scalar field space and will remain constant in time through the bounce until the conversion phase, and its perturbations will obtain a nearly scale invariant spectrum. This is due to the function $\Omega^2(\phi)$, which we describe below, but during ekyprosis $\chi = const$, so this term is irrelevant until the conversion phase. Entropy perturbations, defined using the perturbations of both of these scalar fields, will evolve and then source curvature perturbations during the conversion phase- induced by a bend in the field space trajectory- which will result in the observable quantities in the CMB.


In the following subsections we will define the form of the potential and other terms in the action, and then discuss the two classes of models of interest, defined by whether the conversion of entropy to curvature modes occurs before or after the bounce. We will then discuss the equations of motion of the perturbations, and how we obtain our predictions for the cosmological parameters from these perturbations.

\subsection{The Ekpyrotic and bounce phases}

An Ekpyrotic phase occurs when the energy density is dominated by a potential of the form
\begin{align} V(\phi) = -V_o e^{\sqrt{2\epsilon}\phi} \end{align}
for $\phi < 0$, with $\epsilon = \frac{3}{2}(1+w)$, where $\phi$ has equation of state $w \equiv \frac{P}{\rho} >> 1$, for density $\rho$ and pressure $P$. This potential is steep and negative, in contrast with the flat and positive potential of most simple inflation models. Such a phase dominated by this high pressure scalar field can solve the flatness and homogeneity problems (as does inflation) \citep{ek, review} as well as providing a mechanism for the seeds of cosmic structure formation.

Below is a sketch of an ekpyrotic potential. Details are model-dependent, but in any case Ekpyrosis is defined by a steep, negative potential: 
\begin{figure}[H]
    \centering
    \includegraphics[width=.7\linewidth]{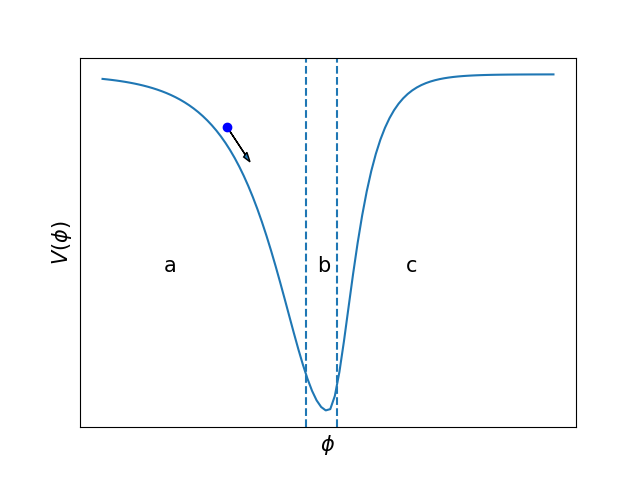}
\end{figure}
In region $(a)$, the field rolls quickly to the right toward the minimum during Ekpyrosis, resulting in a slow contraction of the scale factor. In $(b)$, Ekpyrosis comes to an end near the minimum, and in $(c)$ the potential becomes negligible and at some point the scale factor starts to expand again. 

The kinetic term in the action (2) is $P(X,\phi) = K(\phi)X + Q(\phi)X^2$, where $X \equiv \frac{1}{2}\partial_{\mu} \phi \partial^{\mu} \phi$, and allows for a nonsingular bounce. It reduces to the canonical kinetic energy $P(X,\phi) \approx X$ for most of cosmic evolution. These functions are
\begin{align} K(\phi) = 1 - \frac{2}{(1+\frac{1}{2}\phi^2)^2}, \quad Q(\phi) = \frac{q}{(1+\frac{1}{2}\phi^2)^2} \end{align}
and they allow for a period of null-energy condition violation- and hence a bounce- in the interval $t \in (-\sqrt{q},\sqrt{q})$. The field $\phi$ is a ghost condensate, meaning a scalar field with wrong-sign kinetic energy to violate the null-energy condition. Excitations of the field then have energy unbounded from below, so the term $Q(\phi)X^2$ is added to prevent catastrophic vacuum instabilities \citep{ghost, nec}. 

In the following, we define the two classes of models - in which the curvature modes are generated after or before a bounce - in terms of the potentials an kinetic coupling functions $\Omega(\phi)$. Then, we describe the mean-field solutions and then the perturbative solutions. 
\subsection{Postbounce Conversion models}
In the first two models, a bounce is caused by the function $P(X,\phi)$ and then a conversion of entropy to curvature modes occurs afterward. There is a ``Stable" and an ``Unstable" model in this situation. These are differentiated by the form of the potential and the presence or absence of a kinetic coupling term between $\phi$ and $\chi$,
\begin{align}Stable: \quad V(\phi) = -\frac{2V_o}{e^{-\sqrt{2\epsilon}\phi} + e^{\sqrt{2\epsilon}\phi}} \quad , \quad \Omega^2(\phi) = e^{-b\phi}  \end{align}
\begin{align}Unstable: \quad V(\phi,\chi) = -\frac{2V_o}{e^{-\sqrt{2\epsilon}\phi} + e^{\sqrt{2\epsilon}\phi}}(1 + \frac{1}{2}\epsilon \chi^2),  \quad \Omega^2(\phi) = 1  \end{align}
where the potential is the same as (3) (up to a factor of $2$) for large negative $\phi$. In each case, the full potential is the Ekpyrotic one plus a repulsive potential
\begin{align}V_{rep} = 10^{-4}V_o exp\left(-5[sin(\pi/10)\phi-cos(\pi/10)\chi+2]^2\right) \end{align}
which is a smooth, Gaussian-shaped barrier oriented at an angle to the field space trajectory. This leads to a conversion phase after the bounce via a bend in the field space trajectory. We note the latter case requires special initial conditions (hence the label ``Unstable"), but this may be avoided in the cyclic picture \citep{dark}. 

\subsubsection{Mean-field solution}
When solving the equations of motion, the Lagrangian is broken up into the zeroth-order background solution (describing the energy density and Hubble dynamics), and the perturbations to this background. The first order perturbations describe the linear growth of structure and higher orders describe coupling between modes. Beginning with the background equations, first there is the Friedmann equation
\begin{align} 3H^2 = \frac{1}{2}K\dot{\phi}^2+\frac{3}{4}Q\dot{\phi}^4+\frac{1}{2}\dot{\chi}^2+V \end{align}
which is equal to the total density. Then, the equation of motion for $\phi$ which drives the bounce and Ekpyrotic phases is
\begin{align}\ddot{\phi} = \frac{ \frac{1}{2}K'\dot{\phi}^2+\frac{1}{4}Q'\dot{\phi}^4-V' -3H\dot{\phi}(K+Q\dot{\phi}^2)-(K'+Q'\dot{\phi}^2)\dot{\phi}^2 }{K+3Q\dot{\phi}^2} \end{align}
where primes denote derivatives with respect to $\phi$. Away from the bounce, this reduces to the usual Klein-Gordon equation $\ddot{\phi}+3H\dot{\phi} +V_{\phi} = 0$. This simpler case is the form for $\chi$,
\begin{align} \ddot{\chi} + 3H\dot{\chi}+V_{,\chi} = 0. \end{align}

After defining the potential and its derivatives, we start at $t = 0$ and choose $\phi = 0$ and  require $\rho = 0$ (for the bounce to occur at this time). To enforce the latter, we need
\begin{align} \dot{\phi}(t=0) = -\sqrt{\frac{1+(1+12qV_o)^{1/2}}{3q}}. \end{align}
We can then evolve the background fields to some desired (negative) initial time.

We then use the resulting $\phi,\dot{\phi}$ to define $H(t)$, which is also gives $N(t)$ and then $a(t)$ via
\begin{align} N = \int H dt \quad , \quad a = e^N.\end{align} 
Integrating the inverse scale factor gives the initial conformal time $\tau$, which is needed for the Fourier mode of the entropy perturbation at some initial time, which we will need to obtain the amplitude of the power spectrum at the end of evolution (see $\S 2.4$). 

\subsubsection{Perturbations}
Now we consider the perturbations up to second order. There is the entropy perturbation $\delta s \equiv \frac{-\dot{\chi}\delta \phi+\dot{\phi}\delta \chi}{\sqrt{\dot{\phi}^2+\dot{\chi}^2}}$  which acquires a scale-invariant spectrum and the conversion to curvature perturbations $\zeta$, which is used in the standard definition of the primordial power spectrum and is linearly related to the density perturbation. The evolution of these perturbations is given by

\begin{eqnarray}
  \ddot{\delta s} & =-3H\dot{\delta s}-(V_{ss}+3\dot{\theta}^2)\delta s 
          -\frac{\dot{\theta}}{\dot{\sigma}}(\dot{\delta s})^2-\frac{2}{\dot{\sigma}}\left(\ddot{\theta}+V_{\sigma}\frac{\dot{\theta}}{\dot{\sigma}}-\frac{3}{2}H\dot{\theta}\right)\delta s \dot{\delta s} \\
          & +\left(-\frac{1}{2}V_{sss}+5V_{ss}\frac{\dot{\theta}}{\dot{\sigma}}+9\frac{\dot{\theta}^3}{\dot{\sigma}}\right)(\delta s)^2 \nonumber
\end{eqnarray}

\begin{align} \dot{\zeta} = -\frac{2H}{\dot{\sigma}}\dot{\theta}\delta s + \frac{H}{\dot{\sigma}^2}(V_{ss}+4\dot{\theta}^2)(\delta s)^2-\frac{V_{\sigma}}{\dot{\sigma}}\delta s \dot{\delta s}. \end{align}
Using eqs. (8-10) and (13-14), we can evolve the background and perturbations through the bounce until the end of the conversion phase.



\subsection{Prebounce conversion models}
The other class of models we consider has a conversion phase before the bounce \citep{nonmin}. The theory is still defined by the action in eq. (2), but the kinetic term is $P(\phi,X) = \frac{1}{2}\partial_{\mu} \phi \partial^{\mu} \phi$. Details of the bounce are not considered in these cases, and it is assumed modes pass through the bounce unchanged. The effective form of the coupling function $\Omega^2(\phi)$ changes between the Ekpyrotic and conversion phases. This coupling takes the same form $\Omega^2 = e^{-b\phi}$ during the Ekpyrotic phase, but during the conversion, the two forms considered by \citep{nonmin} are
\begin{align} \Omega^{Bessel}(\phi) = 1 - b I_o(d e^{c \phi/2}) \end{align}
\begin{align} \Omega^{Asymptotic}(\phi) = 1 - b e^{d \phi/2}\end{align}
where $I_o$ is a modified Bessful function of the first kind. 

The potential is the Ekpyrotic one (negligibly small during the conversion phase) plus one of the four repulsive potentials defining the model
\begin{align}V^{rep}_{1,2} = v [x^{-2} + r x^{-6}], v[(sinh x)^{-2} + r(sinh x)^{-4}]  \end{align}
with $r=0,1$ and $x \equiv -\frac{\phi}{2}+\frac{\sqrt{3}\chi}{2}$. Solving the equations of motion in this case is simpler than for the previous two models because we simply start at an initial time (the end of the Ekpyrotic phase) and evolve forward to some time near the bounce at $t=0$. 

\subsubsection{Mean-field and perturbative solution}
The equations of motion are similar to eqs (8-10), but simpler without the inclusion of the bounce:
\begin{align} 3H^2 = \frac{1}{2}\dot{\phi}^2+\frac{1}{2}\Omega^2(\phi)\dot{\chi}^2 +V(\phi) \end{align}
\begin{align} \ddot{\phi} + 3H\dot{\phi}+V_{,\phi}-\Omega\Omega_{,\phi}\dot{\chi}^2 = 0 \end{align}
\begin{align} \ddot{\chi} + (3H+2\frac{\Omega_{,\phi}}{\Omega}\dot{\phi})\dot{\chi}+\Omega^{-2}V_{,\chi} = 0 \end{align}

The perturbations, however, are a bit more complicated here due to a different definition of $\delta s$,
\begin{eqnarray}
  \ddot{\delta s}&=&-3H\dot{\delta s}-(V_{ss}+3\dot{\theta}^2 +\dot{\sigma}^2 e_{s}^{I}e_{s}^{J}e_{\sigma}^{K}e_{\sigma}^{L}R_{IJKL})\delta s 
          - \frac{\dot{\theta}}{\dot{\sigma}}(\dot{\delta s})^2-\frac{2}{\dot{\sigma}}\left(\ddot{\theta}+V_{\sigma}\frac{\dot{\theta}}{\dot{\sigma}}-\frac{3}{2}H\dot{\theta}\right)\delta s \dot{\delta s} \nonumber \\
          &+&\left(-\frac{1}{2}V_{sss}+5V_{ss}\frac{\dot{\theta}}{\dot{\sigma}}+9\frac{\dot{\theta}^3}{\dot{\sigma}} 
          + e_{s}^{I}e_{s}^{J}e_{\sigma}^{K}e_{\sigma}^{L}\left(\dot{\sigma}\dot{\theta}R_{IKJL}-\frac{1}{2}\dot{\sigma}^2e_s^N \mathcal{D}_N R_{IKJL} \right)\right)(\delta s)^2
\end{eqnarray}
where $e_{\sigma}^I = (\frac{\dot{\phi}}{\dot{\sigma}},\frac{\dot{\chi}}{\dot{\sigma}}), e_s^I = (-\Omega \frac{\dot{\chi}}{\dot{\sigma}}, \frac{\dot{\phi}}{\Omega\dot{\sigma}})$ are unit vectors in field space and $R^I_{JKL}$ is the Riemann tensor associated with the field space metric $G_{IJ} = diag(1,\Omega^2)$. The evolution of $\zeta$ is the same as in eq (14).

\subsection*{Summarizing the types of models}
We are considering ten Ekpyrotic models:

\begin{enumerate}
  
  \item [1.] Stable potential, ``standard'' kinetic coupling $\Omega = e^{-\frac{b}{2}\phi}$, conversion after bounce

 \item [2.] Unstable potential, no kinetic coupling, conversion after bounce

 \item [3-6.] Stable potential, Bessel-type kinetic coupling, conversion before bounce

 \item [7-10.] Stable potential, asymptotic-type kinetic coupling, conversion before bounce
\end{enumerate}

where 3-6 and 7-10 are differentiated by the four different possibilities for the repulsive potential, eq (17).

These models have been presented as a proof-of-concept \citep{after, nonmin}, showing that Ekpyrotic theory, with specified Ekpyrotic potentials, repulsive potentials, and field space couplings, can give results consistent with modern CMB data. We extend this in two ways. First, by exploring their model parameter spaces to be able to quantify how often these parameters give results for observables consistent with Planck, and second by numerically calculating the amplitude of the power spectrum in addition to the bispectrum and trispectrum parameters.
\subsection{Observables from the Power spectrum}
We can now relate the cosmological observables to the quantities we calculate from the above equations of motion. The dimensionless primordial power spectrum is often written
\begin{align} \Delta_{\zeta}^2 = \frac{k^3}{2\pi^2}|\zeta_k|^2 = A_S \left( \frac{k}{k_*}\right)^{n_S - 1} \end{align}
where $\zeta_k$ is the Fourier Transform of $\zeta(x)$.

The spectral index of scalar perturbations $n_S$ and its running $\alpha_S \equiv \frac{d n_S}{d lnk}$ can be determined analytically, unlike the other observables, and are set well before the bounce during the Ekpyrotic phase. To do this, we start from the linear equation of motion of the entropy field. For the minimal model (and using the rescaled variable $v_k = a \delta s_k$), this is
\begin{align}v_k'' + (k^2 - \frac{a''}{a}+a^2 V_{ss})v_k = 0  \end{align}
and for the nonminimal ones,
\begin{align}v_k'' + \left(k^2 - \frac{a''}{a}+\frac{\Omega_{,\phi}}{\Omega}a^2V_{,\phi} - \frac{\Omega_{,\phi \phi}}{\Omega}(\phi')^2\right)v_k = 0 . \end{align}
The solution in either case is 
\begin{align} v_k = \frac{\sqrt{-\pi \tau}}{2} H_{\nu}^{(1)}(-k\tau) \end{align}
where $H_{\nu}^{(1)}$ is the Hankel function of the first kind. The index $\nu$ depends on the terms in parenthesis in each of the equations of motion. 

From this definition,
\begin{align} n_S = 4 - 2\nu. \end{align}

The running of the spectral index can then be found using $\alpha_S \equiv \frac{d n_S}{d lnk} = \frac{d nS}{d\mathcal{N}}$, where $d\mathcal{N} = d ln(a|H|)$ and $\mathcal{N}$ is the number of e-folds before the end of Ekpyrosis. The results are in Table 1 \citep{running}.

\begin{table}[h!]
\centering
 \begin{tabular}{||c @{\hskip 1in} p{8cm} ||}
 \hline
 $\Omega^2(\phi) = e^{-b\phi}$ & $\Omega^2(\phi) = 1$ \\ [0.5ex] 
 \hline\hline
 $n_S = 1 - \frac{2\epsilon}{\epsilon - 1}\left(\frac{b}{\sqrt{2\epsilon}}-1\right) - \frac{7}{3}\frac{d ln\epsilon}{d\mathcal{N}}$  & $n_S = 1+\frac{2}{\epsilon} - \frac{\epsilon_{,\mathcal{N}}}{\epsilon}$  \\ 
 $\alpha_S = -\frac{b}{\sqrt{2\epsilon}}\frac{dln\epsilon}{d\mathcal{N}} +\frac{7}{3}\frac{d^2ln\epsilon}{d\mathcal{N}^2}$  & $\alpha_S = \frac{2\epsilon_{,\mathcal{N}}}{\epsilon^2} + \frac{\epsilon_{,\mathcal{N}\mathcal{N}}}{\epsilon} -(\frac{\epsilon_{,\mathcal{N}}}{\epsilon})^2$  \\ [1ex] 
 \hline
 \end{tabular}
 \caption{\label{table-name} Spectral index and its running for each type of kinetic coupling. The right column is applicable only to model 2, the ``Unstable" model.}
\end{table}

Finally, we get the amplitude of the spectrum
\begin{align} A_S = \frac{\zeta(x,t_f)^2}{2\pi^2}\left(\frac{1}{-\tau_f a \sqrt{2} \delta s_i}\right)^2 \left(\frac{-\tau_f k}{2}\right)^{n_S-1} \end{align}
where the real space perturbation $\zeta(x,t_f)$ is the result of solving the equations of motion and $\delta s_i$ is the initial value for the entropy perturbation. We rescale with $\zeta_k = \zeta(x,t_f) \frac{\delta s_{k,i}}{\delta s(x)_i}$ and use the approximation of the Hankel function for small argument $H_{\nu}(-k\tau) \approx \frac{1}{\pi}\frac{\sqrt{\pi}}{2}(\frac{2}{-k\tau})^{\nu}$. We use $k = 10^{-30}$, corresponding to modes that left the horizon around $50-60$ efolds before the end of Ekpyrosis (see eg \citep{effective}).

To calculate the spectral index and its running, we must parameterize the equation of state $\epsilon = \epsilon(\mathcal{N})$. Of the ten models, only one does not have a non-minimal kinetic coupling, and so this one (Postbounce conversion, ``Unstable") has a different form of equation of state and spectral index. In all cases we require $n_S = n_S^{Planck}$, the best-fit value from Planck. We do this for ease of computation, as we now have a $4D$ parameter space.

In all cases, of all the model parameters, the Ekpyrotic equation of state depends only on $n$, where $\epsilon \sim \mathcal{N}^n$ and we evaluate at $\mathcal{N} = 60$ e-folds before end of Ekpyrosis at horizon exit (roughly what is needed to solve the flatness and horizon problems). In the two postbounce models, $\epsilon$ has a different value in the kinetic and bounce phases that follow Ekpyrosis, which we treat as a separate parameter.

In the nonminimal cases, the equation of state can be written  \citep{scale}
\begin{align} \epsilon_{ek} = 3 (\mathcal{N} + 1)^n \end{align}
and, for each $n$ sampled, we fix $b$ to fix the spectral index. For the postbounce ``Stable" model, even though Ekpyrosis ends when $\mathcal{N} = 0$, we use different fixed values for $\epsilon$ (that is, not $3$) for the post-Ekpyrotic numerical evolution. Eq (28) is just a useful form during the Ekpyrotic phase in the far past.


In the minimal case,
\begin{align} \epsilon_{ek} = \epsilon_o (\mathcal{N} + 1)^n \end{align}
where $\epsilon_o$ is fixed to fix the spectral index, for each $n$.

\subsection{Non-gaussianity}
Beyond the power spectrum, if the primordial density perturbations are not exactly Gaussian-distributed, then the bispectrum and perhaps trispectrum of perturbations (related to the three- and four-point functions) are needed also to characterize the perturbations. These Ekpyrotic models only generate nongaussianity of the local type \citep{review}, and so we can parameterize it via
\begin{align} \zeta(x) = \zeta_L + \frac{3}{5}f_{NL}\zeta_L^2 + \frac{9}{25}g_{NL}\zeta_L^3  \end{align}
where $\zeta(x) = \zeta_L + \zeta^{(2)} + \zeta^{(3)}$ is the total curvature perturbation, and $\zeta_L$ is exactly gaussian. Local nongaussianity can be understood as very long wavelength modes affecting the spectrum of shorter modes. For our case, both parameters are exactly zero until the conversion phase begins \citep{nonmin}. The kinetic coupling term in the Prebounce conversion models affects also the non-gaussianity parameters, and not just $n_S$ and $\alpha_S$ as is the case in the Stable Postbounce model. 

So, the two nongaussianity parameters can be found by breaking up the equations of motion for $\delta s$ and $\zeta$ into first, second, and third orders. The first order solution is $\zeta^L$, and by solving the second and third order equations we get the parameters by
\begin{align} f_{NL} = \frac{5}{3} \frac{\zeta^{(2)}(t_{final})}{\left(\zeta^L(t_{final})\right)^2} \end{align}
\begin{align} g_{NL} = \frac{25}{9} \frac{\zeta^{(3)}(t_{final})}{(\zeta^L(t_{final}))^3}. \end{align}
$\zeta$, and therefore $f_{NL}$ and $g_{NL}$, approaches a constant after the conversion phase to all orders.

\section{Calculating the relative model probabilities} 
We perform our Bayesian analysis of early universe model parameters using a 5-dimensional parameter space of primordial universe observables with constraints found by the Planck telescope. These observables describe the large scale power spectrum and higher moments. The $68\%$ confidence intervals on the five parameters from Planck are \citep{planck2018param, planck2018NG}

\begin{center}
\begin{tabular}{ |c| } 
 \hline $\begin{aligned}
	ln(10^{10}A_S) & = 3.049 \pm 0.017 \\
	f_{NL} & = -0.9 \pm 5.1 \\ 
 	g_{NL} & = (-5.8 \pm 6.5) \times 10^4 \\ 
 	n_S & = 0.9635 \pm 0.0046 \\
 	\alpha_S & = -0.0055 \pm 0.0067 
        \end{aligned}$ 
        \\ \hline
\end{tabular}
\end{center}

 In the following subsections, we will give an overview of our code pipeline from model parameters to likelihoods and Bayesian evidences, give more detail on this calculation and the Occam's razor principle, discuss our adaptive sampling technique to do this, and discuss our choice of prior.
 \subsection{Pipeline}
In our parametrization, the parameters for the models are
\begin{enumerate}
\item Postbounce unstable: $(\epsilon, n, Q, ln V_o)$
\item Postbounce stable: $(\epsilon, n, Q, ln V_o)$
\item Prebounce Bessel: $(c,n,d, ln v)$
\item Prebounce asymptotic: $(b, n,d,ln  v)$
\end{enumerate}
all to be constrained by the observables $(\alpha_S,A_S,f_{NL},g_{NL})$, where $n_S$ has already been fixed in each case as discussed. Note also there is a constraint $qV_o < 1/4$, or else the bounce will not occur, so we cannot vary these two parameters arbitrarily. For our calculations, we vary $V_o$ and the parameter $Q \equiv 5qV_o \in (0,5/4)$.

We evolve the perturbations as discussed in the previous section until the end of the conversion phase for every desired combination of model parameters, which outputs the three desired observables $A_S, f_{NL}, g_{NL}$. The other two, $ n_S$ and $\alpha_S$, are calculated analytically. We use these and the Planck constraints for the parameters to define a likelihood as a function of model parameters and integrate using spline interpolation.

We compare the Ekpyrotic models with a representative inflation model. We choose Higgs inflation, a one-parameter single field model which is one of the models preferred by Planck \citep{assess}. This model was originally proposed in terms of a modification to general relativity \citep{starobinsky2, vilenkin}, and the potential can be written in a number of ways. Here, we choose
\begin{align} V(\phi) = M^4 (1-e^{-\sqrt{2/3} \phi})^2 \end{align}
which is a limiting case of the $\alpha$-attractor class of inflation models \citep{alpha}. For an example of a Bayesian analysis of this type of potential in the context of dark energy, see \citep{alphade}. This potential has only one free parameter $M$, which determines $A_S$. The other four observational parameters can be found analytically from the form of the potential.

\subsection{Bayesian analysis}
Bayes' theorem can be applied at multiple levels. At the first, parameter estimation, the posterior distribution for a model's parameters $\theta_i$ is
\begin{align} p(\theta_i|M,D) = \frac{\pi(\theta_i)\mathcal{L}(\theta_i)}{\int \pi(\theta_i)\mathcal{L}(\theta_i) d\theta_i} \end{align}
where the prior $\pi(\theta_i)$ must be specified based on some reasonable physical constraints and the likelihood $\mathcal{L}(\theta_i)$ is  constructed from the data and predictions from the model. The evidence $E \equiv \int \pi(\theta_i)\mathcal{L}(\theta_i) d\theta_i$ is just a normalization here (but will be central when performing model comparison).

From the Planck covariance matrix $\sigma_{ij}$ of observables and the values of those observables predicted by an Ekpyrotic model $(A_S, n_S,\alpha_S, f_{NL}, g_{NL})$, we can derive the corresponding $\chi^2(\theta_i)$ and likelihood function $\mathcal{L}(\theta_i) = e^{-\frac{1}{2}\chi^2(\theta_i)}$. We can determine best fit values of model parameters by minimizing $\chi^2$ with respect to these parameters.  With $\vec{y}$ as the vector of observables, $\chi^2$ is 
\begin{align} \chi^2 = (\Delta y_i ) (\Delta y_j )(\sigma^{-1})^{ij} \end{align}
where $\Delta y =\vec{y}_{data}-\vec{y}_{theory}$ is the residual of the vector of observables, for a given set of input model parameters.
The maximum of the posterior gives the most probable parameters $\theta^{MP}$.

At the next level- model comparison-, the evidence from the level of parameter estimation becomes the likelihood for a model. Bayes theorem can here be written
\begin{align} p(M|D) = \frac{p(M)p(D|M)}{p(D)} \end{align}
where $p(M)$ is the model prior and the model likelihood $\big($which can be written schematically as $p(D|M) = \int p(D| \theta_i M)p(\theta_i|M) d\theta$ $\big)$ is the Evidence $E$ defined above.

We can now illustrate why Bayesian model comparison is useful in comparing even models with different numbers of parameters. One can imagine a model with any number of parameters, but all else being equal, we should favor simplicity. It can be easier to achieve a large value for the maximum likelihood if a model has many parameters, but this is a problem of fine-tuning. A very fine-tuned theory is one for which the likelihood has a very narrow peak relative to the prior range, and this is not preferable. It is in this way that we incorporate Occam's razor into the prior choice \citep{mackay}: a model that is simple (fewer free parameters) will lead to a more precise range of predictions compared with one that is complicated (more free parameters). Often the evidence (model likelihood) can be approximated
\begin{align} p(D|M) \approx p(D|\theta_i^{MP}, M) p(\theta_i^{MP}|M)\sigma_{\theta|D} \end{align}
The first factor on the right is the best fit likelihood, and the next two, which can be expressed as $\frac{\sigma_{\theta|D}}{\sigma_{\theta}}$ (the ratio of the likelihood width over the prior width) for a single parameter theory, is sometimes called the Occam factor \citep{mackay}. It is less than one, and penalizes a model for having free parameters and for having a large prior range. This factor describes by how much the prior space collapses into the posterior space.

Moving on, for two models $M_1$ and $M_2$, the first being simpler, from eq (41) the ratio of posteriors will be
\begin{align} \frac{p(M_1|D)}{p(M_2|D)} = \frac{p(M_1)}{p(M_2)} \frac{p(D|M_1)}{p(D|M_2)}. \end{align}
The priors $p(M)$ could in principle be different, that is we could choose to favor one based on aesthetic or subjective grounds, but this is not necessary because the ratio of likelihoods encode the Occam's razor which prefers simpler models. We will use non-committal priors for the models, $p(M_i) = 1/N_{model}$, meaning each is considered equally likely before considering the data.


The Bayesian framework gives us a systematic way to find the most likely values of a model's parameters and to compare the relative merit of different models. It rewards models in proportion to how well they predict the data, in our case the Planck results for primordial parameters, and punishes them for wasted parameter space. The probabilities we obtain are interpreted as degrees of certainty and describe our state of knowledge.

Bayes theorem can be applied at three different levels: parameter estimation, model comparison, and paradigm comparison. The normalization at one level, the evidence, is the likelihood for the next level. We have discussed the first two, and will leave the last one until the next section when comparing Ekpyrosis with an inflation model.

\subsection{Integration and adaptive sampling}
We now discuss how we calculate the evidence, the principle quantity in model comparison, for a range of model parameters. For a uniform prior, the evidence, or marginal likelihood, can be written
\begin{align} E \equiv \int \mathcal{L} \pi(\theta) d\theta \approx \frac{1}{\Delta \theta} \int_{\in \Delta \theta} \mathcal{L} d\theta, \quad \pi & = 
    \begin{cases}
       \frac{1}{\Delta \theta},& \theta_{Min} < \theta < \theta_{Max},\\
       0,& else
    \end{cases} \nonumber \\ \end{align}
where $\theta$ is the vector of model parameters. That is, the samples of the parameter values $\vec{\theta}$ are drawn from a prior $\pi(\vec{\theta})$. The likelihood is negligibly small for most of parameter space, and so it can be computationally expensive to finely sample over the entire prior volume $\Delta \vec{\theta}$, so we do the integration over a smaller volume $\Delta \vec{\theta}'$ which contains the non-negligible likelihood. We adaptively sample this volume by starting with an $8^4$ grid (since there are $4$ model parameters in each case), and refining by looking at neighboring grid cells, but only sampling if the previous likelihood is above some threshold. Putting it together, we calculate
\begin{align} E' \equiv \int \mathcal{L} \pi' d\theta \approx \frac{1}{\Delta \theta'} \sum_i \int_{\in \Delta \theta_i'} \mathcal{L} d\theta , \quad \pi' & = 
    \begin{cases}
       \frac{1}{\Delta \theta'},& \theta_{Min}' < \theta < \theta_{Max}',\\
       0,& else
    \end{cases} \nonumber \\ \end{align}

where $\Delta \vec{\theta}'$ is broken up into cells of size $\Delta \vec{\theta_i}'$ in each of which we use spline interpolation to integrate (though we note a simple summation of the likelihood divided by the number of points gives similar results), and rescale via the parameter space volumes $\Delta \vec{\theta}'$ and $\Delta \vec{\theta}$ (the prior volume) to get 
\begin{align} E = \frac{\Delta \vec{\theta'}}{\Delta \vec{\theta}}E'. \end{align}
where the prior volume is
\begin{align} \Delta \theta =  \Delta^{Prior} R \end{align}
where $\Delta^{Prior}$ is a 4-D cube containing the prior and $R$ is the ratio of points within the cube that are inside the prior over the total number of points. 

So, the algorithm is:

1) To aid computation, find regions in which likelihood is non-negligible (this requires some trial and error, but the analytic parameters $n_S$, $\alpha_S$ help in limiting the parameter space because they are quick to calculate).

2) Within this region, evaluate likelihood on an $8^N$ grid ($N=4$ in our case).

3) Make refinements by creating grids of size $16^N, 32^N$, etc. with these same bounds. For efficiency, we only calculate the likelihood if the nearest value from the previous refinement was above some chosen threshold. 

4) The resulting integration should converge to the true value, and it must be verified that the likelihood falls off to negligibly small values at the borders of this $N$-dimensional rectangle.

Once we have the evidences for a set of models, their performances can be compared with the Jeffreys' scale \citep{sky}, which is logarithmic in the ratios of evidences,
\begin{enumerate}
\item $|ln E_1/E_2| < 1$ : Inconclusive
\item $|ln E_1/E_2| = 1$ :  Weak evidence
\item $|ln E_1/E_2| = 2.5$ : Moderate evidence
\item $|ln E_1/E_2| = 5$ : Strong evidence
\end{enumerate}

\subsection{Prior choice for parameters}
Finally, we require a choice of prior. The prior must be physically realistic and independent of the data used to construct the likelihood function, and results will depend on the chosen prior range.  It should be noted that priors are not a weakness as not having them would give unphysical answers \citep{sky}. The choice of prior is subjective, but from the perspective of Bayesian statistics there can be no inference without assumptions; Bayes' theorem is a procedure for updating degrees of belief, including the different degrees of belief two different people might have, and often prior (external) information is relevant and hence sensible to include. 

We will define our prior by normalizing the ranges for the parameters by requiring that the observable values fall within some reasonable range. Model parameter values are not known a priori, but we do need some bound on them \citep{hunting}. As long as we normalize the prior for all models considered here consistently, the precise range of values of model parameters corresponding to this prior does not matter because all models will then share a common Occam factor penalty. To define our prior, we choose to require $|f_{NL}| < 100, |g_{NL}| < 10^6, 2.85 < ln(10^{10}A_S)< 3.25, |\alpha_S| < 0.1$, all roughly centered around best fits values known previous to Planck 2018. Additionally, when comparing evidences we use the Jeffreys' scale, which is logarithmic in the ratios of evidences, so the dependence of our results on choice of prior is milder than it might seem \citep{best}. 

\section{Results}
\subsection{Ekpyrotic model comparison}
For simplicity, we start by constraining the models using only the nongaussianity parameters $f_{NL}$ and $g_{NL}$. We do this also because it is only these that have been calculated numerically for the asymptotic and Bessel models in \citep{nonmin}. We here only vary two of the parameters for each model. For the unstable model, we vary $\epsilon$ and $Q \equiv 5 q V_o$, fixing $n, lnV_o$. For the stable one, we vary the same fixing $b, ln V_o$. For the Bessel and asymptotic models, we vary $b (= d)$ and $ln v$. With these choices, we find all the evidences to be comparable, with some only weakly favored over others according to the Jeffreys scale, but nothing more definitive than that.  

We now consider also the primordial power spectrum parameters $A_S,n_S,\alpha_S$ to constrain the models. Fig 4 shows the log Bayes factors of the Ekpyrotic models, all compared with a reference model, the Bessel model with $V_{1,r=1}$ (the first form in eq. (17)), which has the highest evidence. For this preferred model, in Fig 3 we show the curvature perturbation. We show $2D$ slices of the likelihood for this reference model in Fig 1. The likelihoods are predominantly determined by $f_{NL}$ and $A_S$, and we show an example demonstrating this in Fig 2. According to the Jeffreys scale, we find that the reference model is moderately preferred over seven of the models and weakly preferred over one, and nothing conclusive can be said about the remaining one. However, the Higgs inflation scenario is moderately preferred over even the best of the Ekpyrotic models here considered.

\subsection{Dimensionality of parameter space}
We can also compare models (with or without different numbers of parameters) by using the  Bayesian Complexity \citep{status, best},
\begin{align} \mathcal{C}^i_b = \langle -2 ln\mathcal{L}(\theta) \rangle + 2 ln \mathcal{L}^{max} \end{align}
a standard part of Bayesian analyses of inflationary models. This is a more sophisticated measure of compelxity than simply the number of parameters and comes from the relative entropy between the prior and posterior. The bracket $<...>$ denotes an average over the posterior. If two models have similar evidences, then $\mathcal{C}^1_b > \mathcal{C}^2_b$ means we ought to prefer model 2, because the data are sufficient to measure the extra parameters of model 1 but that they were not required. Hence, model 2 is simpler.

The effective number of unconstrained parameters is then defined by
\begin{align} N^i_{uc} = N^i - \mathcal{C}^i_b \end{align}
where $N^i$ is the number of model parameters. A small, nonnegative $N^i_{uc}$ is preferred. This is used to distinguish between two models if their evidences are very close. Our results are in Fig 5.

\subsection{Comparison with Higgs inflation}

Now, we look at our one parameter model of single field inflation, defined by the Higgs potential eq (33). A great advantage of single field slow roll inflation is we have fewer parameters to deal with. The form of the potential gives us predictions for some of the observables in a way that is independent of the potential parameters. The mass parameter $M$ in the potential (33) is only constrained by the amplitude of fluctuations $A_S$. The other four observables come from the form of the potential and the value of $\phi$ 60 e-folds before the end of inflation. 

We can get $n_S$ and $\alpha_S$ \citep{tasi}, as well as  $f_{NL}$ and $g_{NL}$ \citep{deltan} analytically to a good approximation, as they depend only on the form of the inflationary potential, and the mass parameter $M$ only affects the prediction for the power spectrum amplitude $A_S$. These first calculations are straightforward and we find $n_S = 0.9678, \alpha_S = -0.0044, f_{NL} = 0.0133, g_{NL }= 1.446 \times 10^{-5}$. The power spectrum is then given by $A_S = \frac{1}{32\pi^2}M^4(1-e^{-\sqrt{2/3}]\phi_*})^2$, where $\phi_*$ corresponds to horizon crossing. We then integrate the resulting likelihood and obtain the evidence using the same prior as for Ekpyrosis. Comparing with  our preferred Ekpyrotic model, we obtain 

\begin{align} ln \frac{E^{B V_{1,r=1}}}{E^{Higgs}} = -3.36 \end{align}
 
 indicating moderate preference of Higgs inflation over our reference Ekpyrotic model. This is also included in Fig 4.


\subsection{Paradigms and Predictivity}

We have discussed two levels at which we can apply Bayes' theorem, parameter estimation and model comparison. The third level is the comparison of different paradigms  \citep{paradigm}, in our case inflation and Ekpyrosis. In reference  \citep{paradigm}, it is pointed out that there are instances when comparing different paradigms, a straightforward noncomittal prior- which we have assumed in eq (45)- may be insufficient. And so we must consider what it means to compare the two, and to prefer one over the other. Here, we must be careful to consider testability of a set of models, in addition to goodness of fit with data. A model or paradigm that is difficult to falsify is always favored by the Bayes factor. We typically do not need to worry about this for model selection because models within a paradigm typically have very similar predictive power. Bayesian evidence alone gives a softer penalty to unpredictivity than to wrong predictions. A model should do well compared with how it would have performed if the data were different.

So, it has been proposed that paradigm priors should be modified accordingly with a measure of predictivity. Predictivity can be defined \citep{paradigm}
\begin{align}P_r = 1-\int_{\bar{E}}^{E_{Max}}P(E)dE \end{align} 
where $E$ is the evidence of either a model or paradigm, and $\bar{E}, E_{Max}$ are a chosen lower bound and the maximum evidence of the model/paradigm. Then we would modify the prior to 
\begin{align}\pi(\mathcal{M}) = \frac{e^{-\frac{(1-P_r)^2}{P_r^2}}}{N} \end{align}
for some normalization $N$. The predictivity here depends on a paradigm's spread of predictions, not the number of models.


So, a successful comparison between the two paradigms will depend on the ratio of posteriors for the two, which will depend on the paradigm evidences, and the predictivities, which depend on how rare it is to get high evidence and our choice for the lower limit $\bar{E}$, which is chosen such that the predictivities don't change very much if $\bar{E}$ is lowered further.

We only consider the simplest inflation models here, and including others (with multiple fields and more parameters) will worsen the paradigm's performance. But the simplest Ekpyrotic models require more parameters than the simplest inflation models, which is a severe disadvantage for the former from a Bayesian viewpoint. However, even if we could conclude that inflation is favored for now, we note that parameter number is ambiguous, because a potential can come from an action (or some fundamental theory) with a different number of parameters \citep{assess}, so if for example the Ekpyrotic bounce can be embedded in a string theory scenario with fewer parameters this will help the theory. 

Following Gubitosi et al, we parameterize the threshold as $\bar{E} = \frac{\bar{E}_{max}}{n e}$, for different values of $n$, and increase $n$ until results stabilize. We look at the predictivity of our representative Ekpyrotic and inflation models using both $f_{NL}$ and $ln(10^{10}A_S)$. Varying $f_{NL}$ only, we find similar predictivities and a posterior ratio not much different from eq (45). Including the amplitude of the power spectrum as well leads to a much different result,
\begin{align}P^{inf}_r = 0.52, P^{ek}_r = 0.079 \end{align}
which modifies the result for the log Bayes factor eq. (45) to the log of the posteriors,
\begin{align} ln\left( \frac{P_{ek}}{P_{inf}} \right) = -138.4 \end{align}
indicating very strong preference for inflation over our reference Ekpyrotic model.


\subsection{Discussion}
We have here done, as far as we know, the first Bayesian analysis of Ekpyrotic models, considering sub-Planck energy level Ekpyrotic models still viable given CMB data from Planck in which entropy perturbations are converted into curvature ones before or after a bounce phase. We have shown that we can discriminate between these models in a Bayesian sense using Planck data, and also that they are all moderately to severely penalized compared with a simple inflation model as a result of the wasted parameter space required. These extra parameters in the Ekpyrotic models are not justified in a Bayesian sense when we have a one-parameter inflation model that can fit the Planck data well enough on its own.

We should stress that we have here chosen one particular parametrization for the Ekpyrotic models. Others could be used instead, for example by instead varying, for instance, the ratio of the Ekpyrotic potential amplitude to that of the repulsive potential in the Postbounce conversion models, which we have not done here. Another possibility is the width of the function $K(\phi)$ parameterizing the width of the bounce, as in \citep{effective}. We note also that different choices of sets of observables result in different evidence distributions, and hence predictivities \citep{paradigm}. We chose to look at nongaussianity in addition to power spectrum parameters, and do not consider the tensor-to-scalar ratio - central to \citep{paradigm}'s analysis - because this will be negligibly small for all Ekpyrotic models. We have also not considered reheating, as the cosmological observables are well approximated in inflation analytically as we have done, and reheating in Ekpyrosis is not nearly as well understood as it is in inflation (and even in inflation's case there is much work to be done).

Nevertheless we have shown, given this particular parameterization, that Ekpyrotic models have an excess of parameters, and while they can achieve a higher likelihood function value, the wasted volume of parameter space results in a disfavorability in a Bayesian sense. This may have been expected given inflation's economy with parameters, but we seen this confirmed here numerically. Improvement on the bounds of the cosmological parameters used for our analysis is expected with future data, and this could potentially alter our results. In the optimistic case, future CMB experiments, for example the proposed Fast Fourier Transform Telescope, can improve on the Planck bound on the running of the spectral index by an order of magnitude \citep{fftt, specrun}. Constraints on the bispectrum and trispectrum probably will not improve with more CMB data, but can with large scale structure data. For example, the Square Kilometer Array is expected to bring the $68\%$ confidence level down by a factor of $5$ \citep{ska}.

\section{Ackowledgements}
We thank Angelika Fertig, Jean-Luc Lehners, and Edward Wilson-Ewing for valuable discussions on Ekpyrotic theory.

\bibliographystyle{unsrt}
\bibliography{PaperDraft}

\begin{figure*}[bp!]
  \begin{subfigure}[t]{0.5\linewidth}
    \centering
     \hspace*{-3cm}
    \includegraphics[width=1\linewidth]{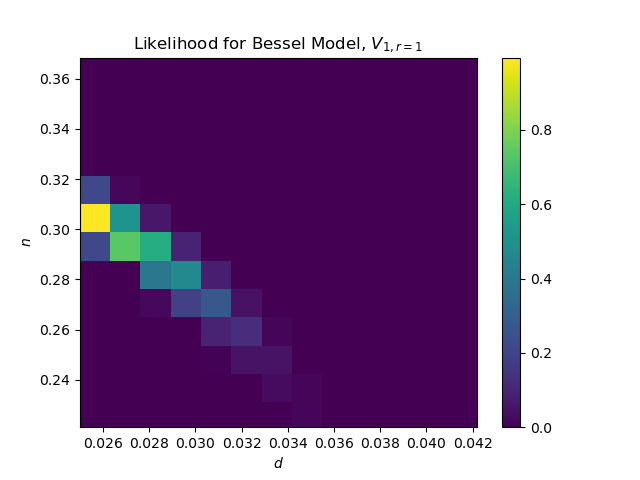} 
    \label{fig7:a} 
  \end{subfigure}
  \begin{subfigure}[t]{0.5\linewidth}
    \centering
    \hspace*{-2cm}
    \includegraphics[width=1\linewidth]{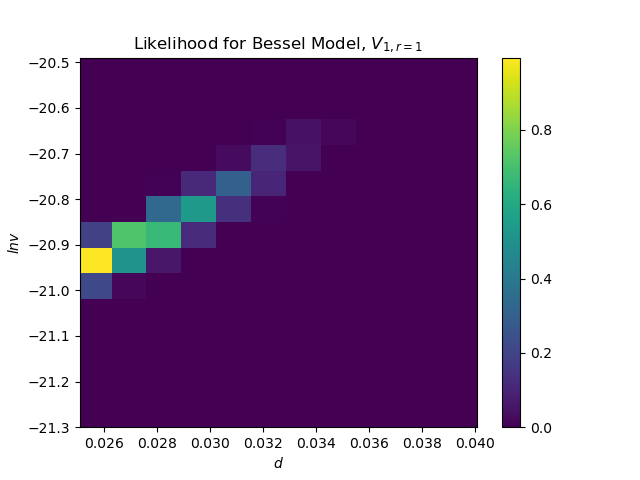}
    \label{fig7:d} 
  \end{subfigure} 
  \begin{subfigure}[t]{0.5\linewidth}
    \centering
    \hspace*{-2cm}
    \includegraphics[width=1\linewidth]{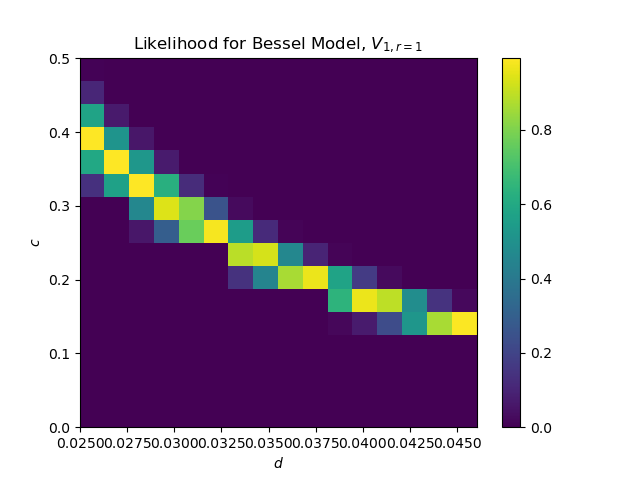}
    \label{fig7:d} 
  \end{subfigure} 
  \begin{subfigure}[t]{0.5\linewidth}
    \centering
    \hspace*{-2cm}
    \includegraphics[width=1\linewidth]{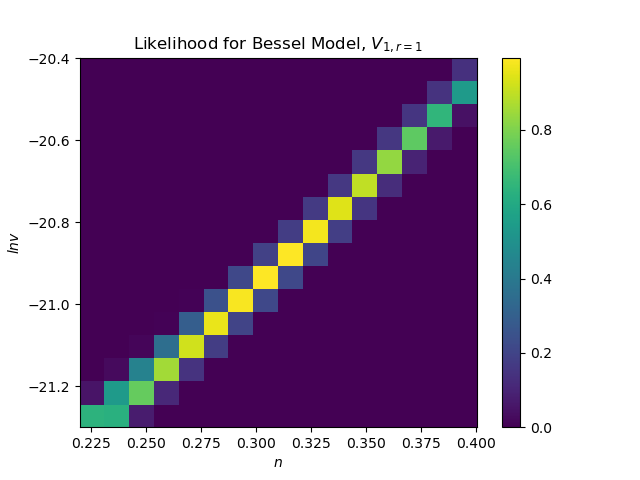}
    \label{fig7:d} 
  \end{subfigure} 
  \begin{subfigure}[t]{0.5\linewidth}
    \centering
    \hspace*{-2cm}
    \includegraphics[width=1\linewidth]{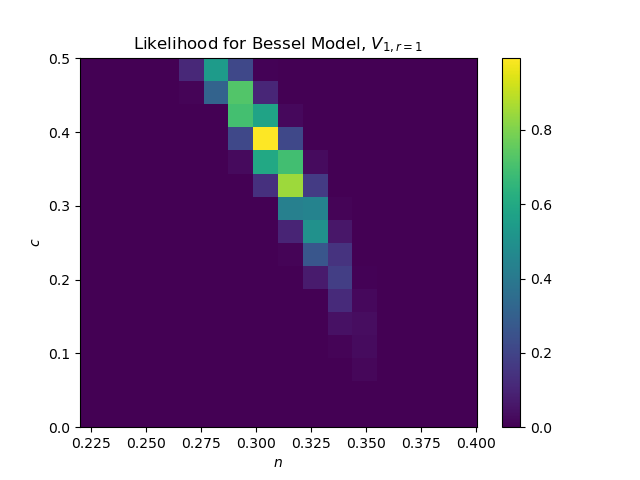}
    \label{fig7:d} 
  \end{subfigure} 
  \begin{subfigure}[t]{0.5\linewidth}
    \centering
    \hspace*{-2cm}
    \includegraphics[width=1\linewidth]{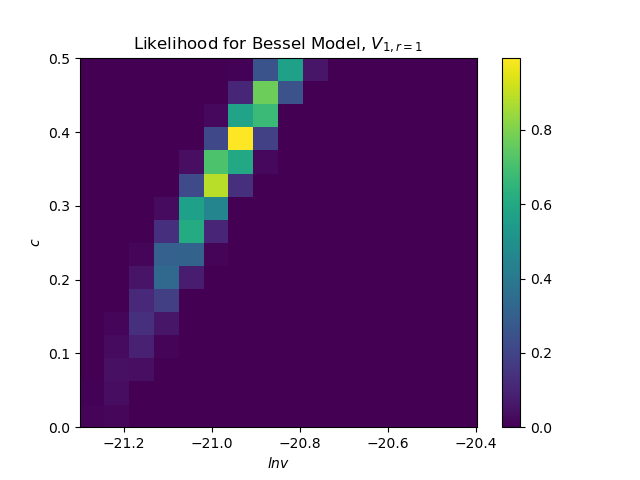}
    \label{fig7:d} 
  \end{subfigure} 
  \caption{2D slices of the likelihood for the best performing Ekpyrotic model considered, with Bessel-type field space metric and potential $V_{1,r=1}$, at the best-fit values of the other two parameters. The model parameters are $(d=0.026, n=0.301, lnv=-20.97,c=.366)$ and the resulting observables are $f_{NL}=-0.64, g_{NL}=-5.9 \times 10^4, ln(10^{10}A_S)=3.0479$.}
  \label{fig7} 
\end{figure*}

\begin{figure*}[bp!]
  \begin{subfigure}[t]{0.65\linewidth}
    \centering
     \hspace*{-3cm}
    \includegraphics[width=1\linewidth]{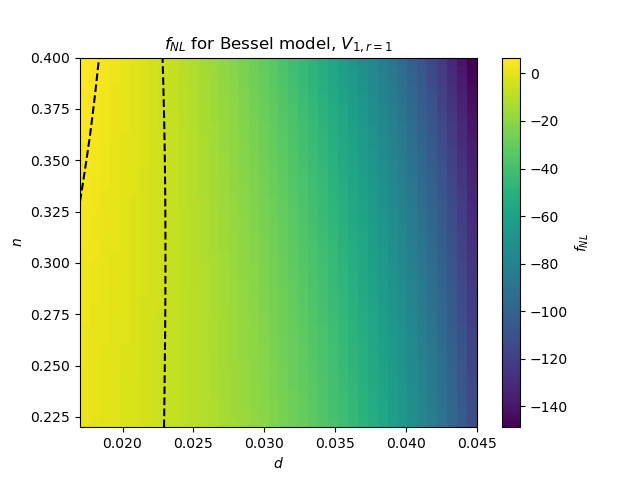} 
    \label{fig7:a} 
  \end{subfigure}
  \begin{subfigure}[t]{0.65\linewidth}
    \centering
    \hspace*{-2cm}
    \includegraphics[width=1\linewidth]{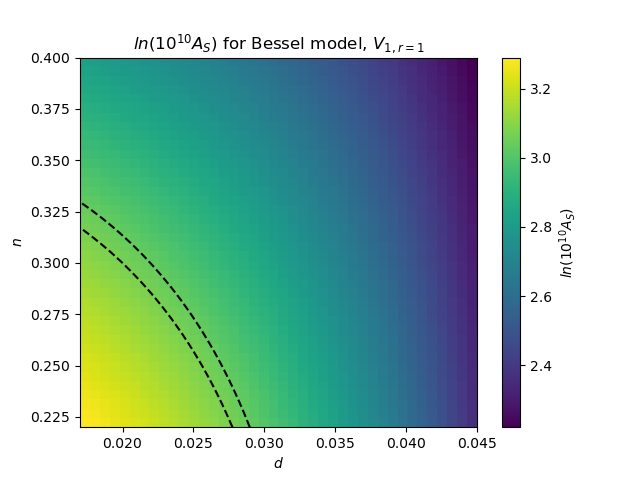}
    \label{fig7:d} 
  \end{subfigure} 
  \caption{Contours for the preferred Bessel-type model, at the best fit values of the other two parameters. Regions within the dotted lines are within $1 \sigma$ of the Planck best fits. Note the overlap of these two $1 \sigma$ regions aligns with the top left plot in Figure 1. }
  \label{fig7} 
\end{figure*}

\begin{figure*}[bp!]
  \begin{subfigure}[t]{0.5\linewidth}
    \centering
     \hspace*{-3cm}
    \includegraphics[width=1\linewidth]{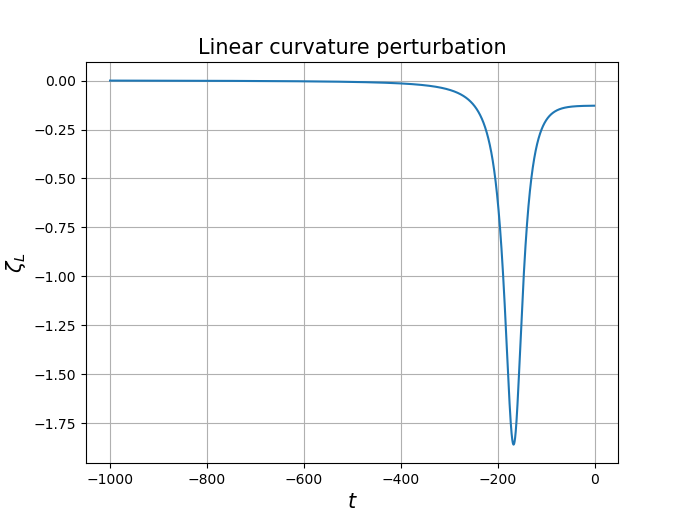} 
    \label{fig7:a} 
  \end{subfigure}
  \begin{subfigure}[t]{0.5\linewidth}
    \centering
    \hspace*{-2cm}
    \includegraphics[width=1\linewidth]{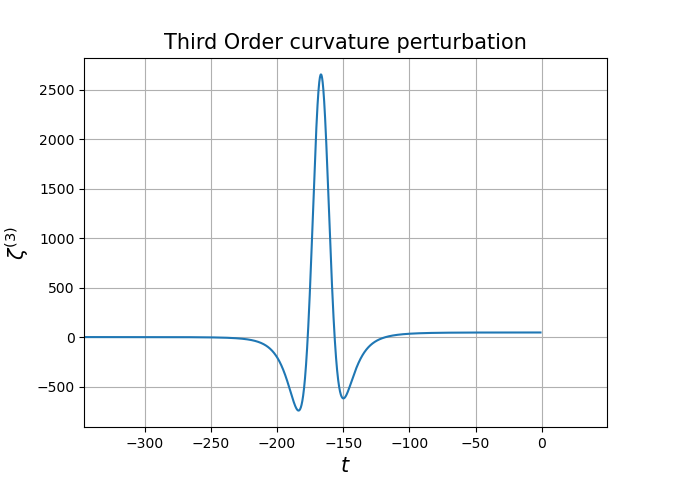}
    \label{fig7:d} 
  \end{subfigure} 
  \begin{subfigure}[t]{0.5\linewidth}
    \centering
    \hspace*{-2cm}
    \includegraphics[width=1\linewidth]{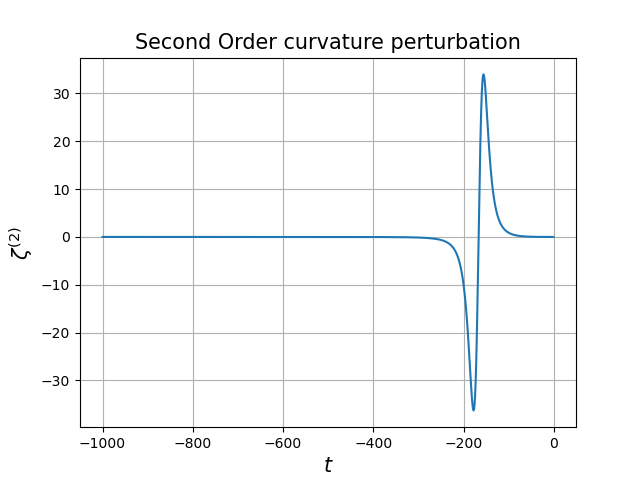}
    \label{fig7:d} 
  \end{subfigure} 
  \begin{subfigure}[t]{0.5\linewidth}
    \centering
    \hspace*{-2cm}
    \includegraphics[width=1\linewidth]{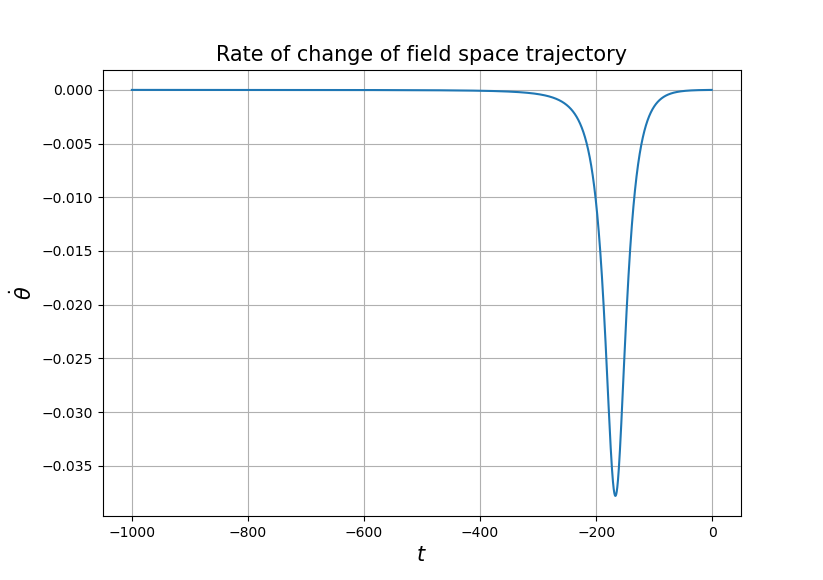}
    \label{fig7:d} 
  \end{subfigure} 
  \caption{Curvature perturbation for model with Bessel-type field space metric and potential $V_{1,r=1}$ with best fit parameters. (Bottom right) The rate of change of the field space trajectory, indicating the start and end of the conversion phase. Initial conditions used are $\delta s^{(1)} = 1, \dot{\delta s^{(1)}} =\frac{\delta s^{(1)} }{t_{end-ek}}$, with all other entropy and curvature perturbations and their time derivatives initially zero. The curvature perturbation is zero until the conversion phase begins, and settles into a constant afterward when the field space trajectory stops turning.}
  \label{fig7} 
\end{figure*}

\begin{figure*}[bp!]
    \includegraphics[width=1\linewidth]{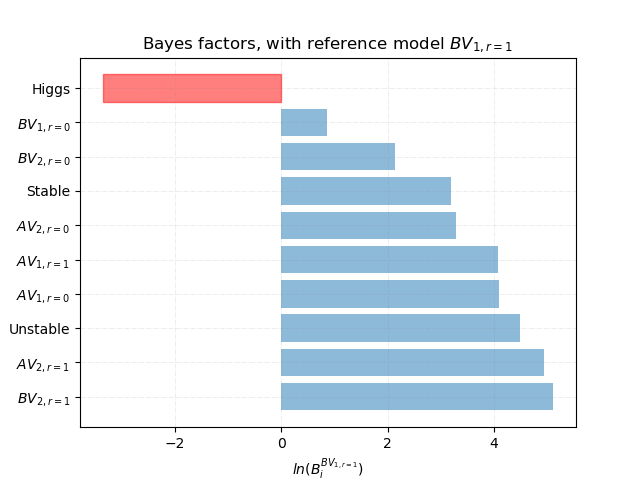}
    \label{fig7:d}  
  \caption{The log of the Bayes factors for the Ekpyrotic models and inflation model considered. The preferred Ekpyrotic model is weakly to moderately favored over eight models  and indistinguishable from one. However, we find Higgs inflation is moderately preferred over even the best Ekpyrotic model on the Jeffreys scale.}
  \label{fig7} 
\end{figure*}

\begin{figure*}[bp!]
    \includegraphics[width=1\linewidth]{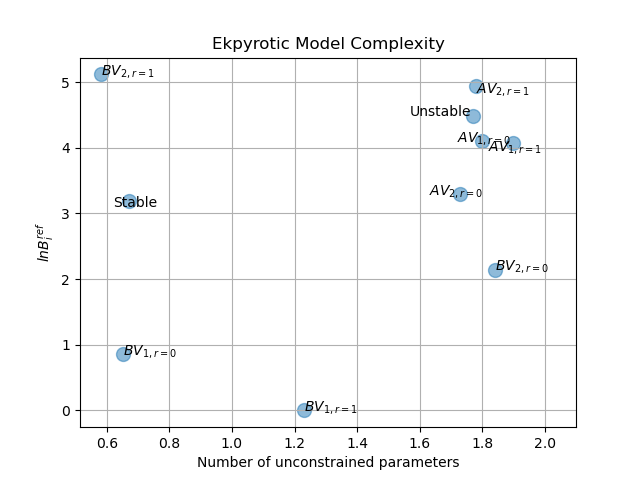}
    \label{fig7:d}  
  \caption{ The log Bayes factor as a function of the effective number of unconstrained parameters. The horizontal direction allows us to distinguish between models with the same evidence, preferring models with the smallest (but nonnegative) $N^{uc}$, with less unconstrained parameters. Along a given horizontal line, the model most to the left is preferred. We note our reference model has $N^{uc} > 1$, but still accept it as our preferred model due to its Bayesian evidence.}
  \label{fig7} 
\end{figure*}




\end{document}